%% file: main.tex
\documentclass[%
 aip,apl,
 amsmath,amssymb,
 reprint,%
]{revtex4-2}

\usepackage{graphicx}
\usepackage{dcolumn}
\usepackage{bm}

\usepackage[utf8]{inputenc}
\usepackage[T1]{fontenc}
\usepackage{mathptmx}
\usepackage{etoolbox}
\usepackage{xcolor}
\usepackage{siunitx}

\DeclareMathOperator{\Det}{Det}
\DeclareMathOperator{\de}{d\!}

\makeatletter
\def\@email#1#2{%
 \endgroup
 \patchcmd{\titleblock@produce}
  {\frontmatter@RRAPformat}
  {\frontmatter@RRAPformat{\produce@RRAP{*#1\href{mailto:#2}{#2}}}\frontmatter@RRAPformat}
  {}{}
}%
\makeatother

\begin{document}

\preprint{AIP/123-QED}

\title{Anomalous supercurrent and diode effect in locally perturbed topological Josephson junctions}
\author{Samuele Fracassi}
\affiliation{Dipartimento di Fisica, Universit\`a di Genova, Via Dodecaneso 33, 16146, Genova, Italy}
\author{Simone Traverso}
\thanks{Author to whom correspondence should be addressed:\\
simone.traverso@edu.unige.it}
\affiliation{Dipartimento di Fisica, Universit\`a di Genova, Via Dodecaneso 33, 16146, Genova, Italy}
\author{Niccolò Traverso Ziani}
\affiliation{Dipartimento di Fisica, Universit\`a di Genova, Via Dodecaneso 33, 16146, Genova, Italy}
\affiliation{CNR-SPIN, Via Dodecaneso 33, 16146, Genova, Italy}
\author{Matteo Carrega}
\affiliation{CNR-SPIN, Via Dodecaneso 33, 16146, Genova, Italy}
\author{Stefan Heun}
\affiliation{NEST, Istituto Nanoscienze-CNR and Scuola Normale Superiore, Piazza San Silvestro 12, 56127 Pisa, Italy}
\author{Maura Sassetti}
\affiliation{Dipartimento di Fisica, Universit\`a di Genova, Via Dodecaneso 33, 16146, Genova, Italy}
\affiliation{CNR-SPIN, Via Dodecaneso 33, 16146, Genova, Italy}

\date{\today}

\begin{abstract}
The simultaneous breaking of time-reversal and inversion symmetry can lead to peculiar effects in Josephson junctions, such as the anomalous Josephson effect or supercurrent rectification, which is a dissipationless analog of the diode effect. Due to their  impact in new quantum technologies, it is important to find robust platforms and external means to manipulate the above effects in a controlled way. Here, we theoretically consider a Josephson junction based on a quantum spin Hall system as the normal channel, subjected to a magnetic field in the direction defined by spin-momentum locking, and in presence of a local tip in close proximity to one of the metallic edges in the normal region. We consider different local perturbations, model normal and magnetic tips, and study how they affect the Josephson response of the device. In particular, we argue that magnetic tips are a useful tool that allows for tunability of both $\phi_0$ response and supercurrent rectification.
\end{abstract}

\maketitle

During the last years, there has been a widespread interest in the physics of low-dimensional semiconductors featuring strong spin-orbit coupling (SOC). Indeed, these systems offer an ideal platform to develop new architectures able to coherently control electron spin, with great impact in spintronics and, through Majorana fermions\cite{mf1,mf2,mf3,mf4,mf5}, topological quantum computing~\cite{Nowack_2007, Bercioux_2015, Manchon_2015, Moehle_2021, Haldane_2017, Prada_2020}. Thanks to their strong SOC, quantum spin Hall systems, such as HgTe/CdTe heterostructures~\cite{Bernevig_2006, Molenkamp_2007, Deacon_2017}, are widely studied platforms presenting, among all, a non-trivial topological phase with conducting helical edge channels\cite{ helical2,helical3,MD1,Dolcetto_2016, Ronetti_2016,helical4,MD2,MD3,helical45,helical5,helical6}. Furthermore, great interest has been recently dedicated to the interplay between strong SOC and superconducting correlations~\cite{Kouwenhoven_2015, Qu_2016, Prada_2020, Rasmussen_2016, Salimian_2021, Iorio_2023, Kaperek_2022, Baumgartner_2022}, leading for instance to several theoretical and experimental studies on topological Josephson junctions (JJs)~\cite{helical1, Deacon_2017, Moehle_2021, Dolcini_2015, Calzona_2022, Blasi_2020, Blasi_2021, Zhang_2022, Vigliotti_2023-I, Vigliotti_2023-II,tj1,tj2,tj3,tj4,SDcitro}. These consist of SNS devices, where the normal (N) part, in between two superconducting (S) contacts, resides in the topological phase. These junctions can display fascinating physics, ranging from the emergence of Majorana bound states~\cite{Prada_2020, helical3} to anomalous dc and ac Josephson responses~\cite{Dolcini_2015, Deacon_2017, helical45, Strambini_2020, Zhang_2022}. In particular, it has been predicted that, in presence of a small magnetic field, a topological JJ can sustain a finite supercurrent at zero phase bias,  the so-called $\phi_0$ effect~\cite{Dolcini_2015}. At the heart of this phenomenon is the simultaneous breaking of time-reversal and inversion symmetry. Notably, it is well known that these two ingredients can lead to non-reciprocal charge transport and form the theoretical basis behind the diode effect. 

Very recently~\cite{Hu_2007, Wakatsuki_2017, Hoshino_2018, Yasuda_2019, Ando_2020, Misaki_2021,tanaka_2022, Daido_2022,burset23, Reinhardt2024, Souto_2024,cayao24}, it has been realized that a superconducting analog of the diode is possible, on the basis of the same arguments, leading to dissipationless supercurrent rectification. The supercurrent diode effect (SDE), can greatly impact low temperature (superconducting) technology.
The first experimental reports on SDE relied  on magnetic or layered bulk materials~\cite{Ando_2020, Noah_2022, Bauriedl_2022, Strambini_2022, Yun_2023, Wu_2022, Sundaresh_2023, Hou_2023}, but recent measurements on JJs, based for instance on  III-V semiconductors, have been also reported~\cite{Turini_2022, Davydova_2022, Souto_2022, Baumgartner_2022, Wang_2024, Pillet_2023}. Still, several questions on dissipationless nonreciprocal transport are open, including its microscopic origin~\cite{Misaki_2021, He_2022, Ilicc_2022, Zhang_2022_b, costa_2023, Steiner_2023, Coraiola_2023b} and its tunability by external means.

In the context of quantum spin Hall systems, in Ref.~\cite{Dolcini_2015} it has been shown by scattering matrix formalism that a bare topological JJ, subjected to a longitudinal (parallel to the direction defined by spin-momentum locking) magnetic field, can give rise to the anomalous Josephson effect. Moreover, finite supercurrent rectification is present along a single helical edge, while the effect disappears when considering the whole sample~\cite{Dolcini_2015}.

Inspired by the success in the field of Wigner crystallization\cite{wt1,wt2,wt3}, where local gating has been proven effective in the manipulation of electrons confined in one dimension, in this work we consider a model of a quantum spin Hall-based topological JJ in the presence of a closely spaced local tip. We model different tips that act as local perturbations on the normal region. First, we show that a tip capacitively coupled to the edge states does not alter at all the supercurrent. Then, we demonstrate that a magnetic tip~\cite{Michlmayr2006, Phark2017, Haze2019} with magnetization parallel to the spin axis defined by spin-momentum locking (here in the $z$-axis) only shifts the current-phase relation (CPR), thus influencing the current at zero phase bias but not the SDE. Finally, a magnetic tip with magnetization perpendicular to the previous one (hence along any direction in the $xy$ plane) influences both the current at zero phase bias and the SDE.

\begin{figure}[t]
    \centering
    \includegraphics[width=\linewidth]{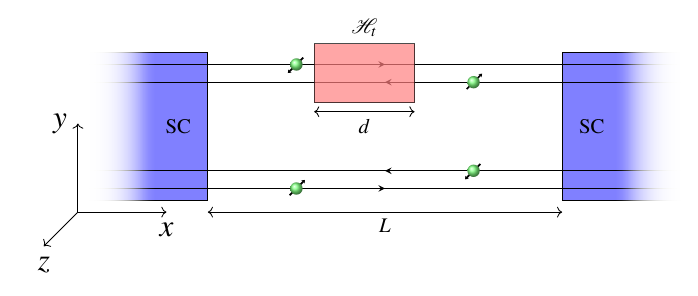}
    \caption{Schematic of the system under study, consisting of a Josephson junction lying in the $xy$ plane, with a topological insulator as the normal part. A tip of size $d$ along the channel axis ($x$-axis) perturbs the upper helical edge states with an Hamiltonian $\mathcal{H}_t$. The direction defined by the spin-momentum locking in assumed parallel to the $z$-axis.}
    \label{fig:setup_scheme}
\end{figure}

The system is modeled as two one-dimensional channels, located at the opposite sides of the structure. Each channel is helical, and the helicity of the two channels is opposite. Moreover, the system is (partially) proximitized by superconducting contacts, so as to define a JJ (see Fig.~\ref{fig:setup_scheme})~\cite{Dolcini_2015}, which can also be subjected to a magnetic field. We also assume that only one of the edges (say the upper one) is perturbed by the tip. Quantitatively, the model we use is represented by the Bogoliubov-de Gennes Hamiltonian~\cite{Beenakker_1991, Beenakker_1992-I, Beenakker_1992-II}
\begin{equation}
    H=\frac{1}{2}\left[\left(\sum_{\lambda=\pm}H_\lambda\right)+H_{t}\right].
\end{equation}
Here, $H_{\lambda}$ is the Hamiltonian of the unperturbed edges, with $\lambda=\pm$ for the upper/lower edge, respectively, and $H_t$   describes the interaction with the tip. We have
\begin{equation}
    H_\lambda=\int_{-\infty}^{\infty}\de x \,\Psi^{\dag}_\lambda(x)\mathcal{H}_\lambda(x)\Psi_\lambda(x),
\end{equation}
with the Hamiltonian density (hereafter we set $\hbar=1$)
\begin{equation}
    \mathcal{H}_\lambda(x)=-\lambda v_F i\partial_x \tau_z\sigma_z+\Delta(x)\tau_x\sigma_0+\mathcal{U}_z\tau_0\sigma_z,
\end{equation}
with basis $\Psi_\lambda(x)=\left(\psi_{\lambda,\uparrow}(x),\psi_{\lambda,\downarrow}(x),\psi^\dag_{\lambda,\downarrow}(x),-\psi^\dag_{\lambda,\uparrow}(x)\right)^T$, where $\psi_{\lambda,\uparrow/\downarrow}(x)$ are the Fermi field operators for electrons in the edge $\lambda$ and with spin projection $\uparrow/\downarrow$.
Here, $v_F$ is the Fermi velocity, $\tau_i$ and $\sigma_i$ ($i=0,x,y,z$) are the identity and the Pauli matrices in particle-hole and spin space respectively and $\Delta(x)=\Delta_0[\theta(-x-L/2)e^{i\varphi/2}+\theta(x-L/2)e^{-i\varphi/2}]$ is the superconducting pairing, with $L$ and $\varphi$ the distance and the phase difference between the two superconducting sections. The parameter $\Delta_0$ represents the effective amplitude of the proximitized superconducting pairing seen by the edge states. Finally, $\mathcal{U}_z=g\mu_BB_z$ is the energy associated to a finite magnetic field $B_z$ via the usual Zeeman coupling, where $\mu_B= \SI{5.788e-5}{\electronvolt\per\tesla}$ is the Bohr magneton and $g$ the effective gyromagnetic factor for the helical edges. Both orbital effects and inter-edge interactions are neglected in our analysis.

The Hamiltonian related to the tip, which only perturbs the upper edge ($\lambda=+$), is given by
\begin{equation}
    H_t= \int_{x_0-d/2}^{x_0+d/2} \de x\sum_{\zeta=1}^3 \Psi^\dagger_+(x) \mathcal{H}_t^{(\zeta)}(x)\Psi_+(x) ,
\end{equation}
where we have imposed that the tip only influences the system for $x_0-d/2<x<x_0+d/2$, with $d>0$ representing the width of the tip, $ \ x_0-d/2>-L/2,\ x_0+d/2<L/2$, and
\begin{align}
    \mathcal{H}_t^{(1)}(x)&=\delta\mu\tau_z\sigma_0,\\
    \mathcal{H}_t^{(2)}(x)&=\delta\mathcal{U}_z\tau_0\sigma_z,\\
    \mathcal{H}_t^{(3)}(x)&=\delta\mathcal{U}_x\tau_z\sigma_x.
\end{align}
Here, $\delta\mu$ represents a local capacitive coupling, and $\delta\mathcal{U}_{z/x} = g\mu_B\delta B_{z/x}$ Zeeman energies related to local magnetic fields parallel or perpendicular to the axis defined by spin-momentum locking, respectively. In the following, the perturbations $\mathcal{H}_t^{(\zeta)}$ will be addressed separately. Notably, none of our results will bear any explicit dependence on the tip position $x_0$. A schematic of the system is presented in Fig.~\ref{fig:setup_scheme}.

The quantity under investigation is the supercurrent $I(\varphi)$ as a function of the phase difference between the two superconducting regions, which are kept at the same voltage. The supercurrent is the sum of the supercurrents $I_\lambda(\varphi)$ carried by the single edges. The current $I_\lambda(\varphi)$ is evaluated by means of the scattering matrix formalism~\cite{Beenakker_1991, Beenakker_1992-I, Beenakker_1992-II}, and in particular by using the relation~\cite{Giazotto_2007}
\begin{equation}
    I_\lambda(\varphi)= -4eT\frac{\partial}{\partial\varphi}\sum_{\nu\,=\,0}^{\infty}\Re\left[\frac{1}{2}\ln [D_\lambda(i\omega_{\nu};\varphi)]\right].
    \label{eq:supercurrent_def}
\end{equation}
Here $\omega_{\nu} = (2\nu+1)\pi T$ are the thermal fermionic Matsubara frequencies and $D_\lambda(z;\varphi)$ is the analytic continuation  of $\Det[\mathbb{I}-S_{\text{bound}_\lambda}(E,\varphi)]$ in the upper complex plane. $S_{\text{bound}_\lambda}$ is the full scattering matrix related to the Andreev bound states. These are subject to the bound state condition $\Psi^{\text{Out}}_\lambda = \Psi^{\text{In}}_\lambda$, where $\Psi^{\text{In}}_\lambda$ describes the \textit{in} states and $\Psi^{\text{Out}}_\lambda$ the \textit{out} states for the edge $\lambda$.~\footnote{The $1/2$ in front of the logarithm is necessary in order not to take into account the negative energy states that arise from particle-hole symmetry.}

For the problem at hand, the scattering matrix $S_{\text{bound}_\lambda}$ is given by $S_{\text{bound}_+}=S_{A_+}S_{\text{prop}_+}$ for the upper (perturbed) edge. In addition, one has $S_{\text{bound}_-}=S_{A_-}S_{\text{prop}_-}^{\text{free}}$ for the lower (unperturbed) edge. 
In the above expressions, $S_{A_\lambda}$ represents the scattering matrix associated to Andreev processes~\cite{Beenakker_1992-I, Beenakker_1992-II}. The explicit form of $S_{\text{bound}_\lambda}$ for the different cases  is given in the Supplementary Material (SM).

As discussed in Ref.~\cite{Dolcini_2015}, in the absence of the tip, one finds that $I_\lambda(\varphi=0)\neq 0$, that is, the single edge shows a $\phi_0$ effect. Moreover, one has that $|\mathrm{Max}_\varphi\left\{I_\lambda(\varphi)\right\}|\neq |\mathrm{Min}_\varphi\left\{I_\lambda(\varphi)\right\}|$, \emph{i.e.},  one has a finite SDE as well. However, one also finds that $I_\lambda(\varphi)=-I_{-\lambda}(-\varphi)$, so that both the $\phi_0$ effect and the SDE vanish for the full system of two edges. Since the rectification in the system without perturbation manifests itself more prominently in the short-junction regime~\cite{Dolcini_2015}, in the following we consider this regime, taking $L/\xi = 0.1$, with $\xi = v_F/\Delta_0$ the superconducting coherence length. For the present work we consider $\Delta_0=\SI{10}{\micro\electronvolt}$~\cite{Kouwenhoven_2015, Deacon_2017} and $v_F=\SI{4.6e4}{\meter\per\second}$, yielding a coherence length of $\xi\sim\SI{3}{\micro\meter}$~\cite{Kouwenhoven_2015, Rui-Rui_2016}. The long junction regime, which can be achieved for instance by increasing $\Delta_0$, is not qualitatively different~\cite{apl1}. However, the effects we discuss are gradually reduced in magnitude as the junction length is increased. 
Furthermore, we choose an effective gyromagnetic factor $g=11.5$ that is a reasonable value for helical edge states~\cite{Rui-Rui_2016}. These values are compatible with those reported for InAs/GaSb~\cite{Kouwenhoven_2015, Rui-Rui_2016}.
Hereafter we also take the tip width $d/L = 0.2$ which corresponds to a reasonable tip physical dimension~\cite{Michlmayr2006, Phark2017, Haze2019, wt1}. 
It is worth anticipating that our results refer to an overall magnetic field $\mathcal{U}_z$ applied in the $z$ direction which is typically considered to be much smaller than the magnetic field induced by the tip. 
Furthermore, we focus on the low temperature regime with $T \ll \Delta_0$, \emph{i.e.}, $T\ll \SI{0.1}{\kelvin}$.

\begin{figure*}[hbt]
    \centering
    \includegraphics[width=\linewidth]{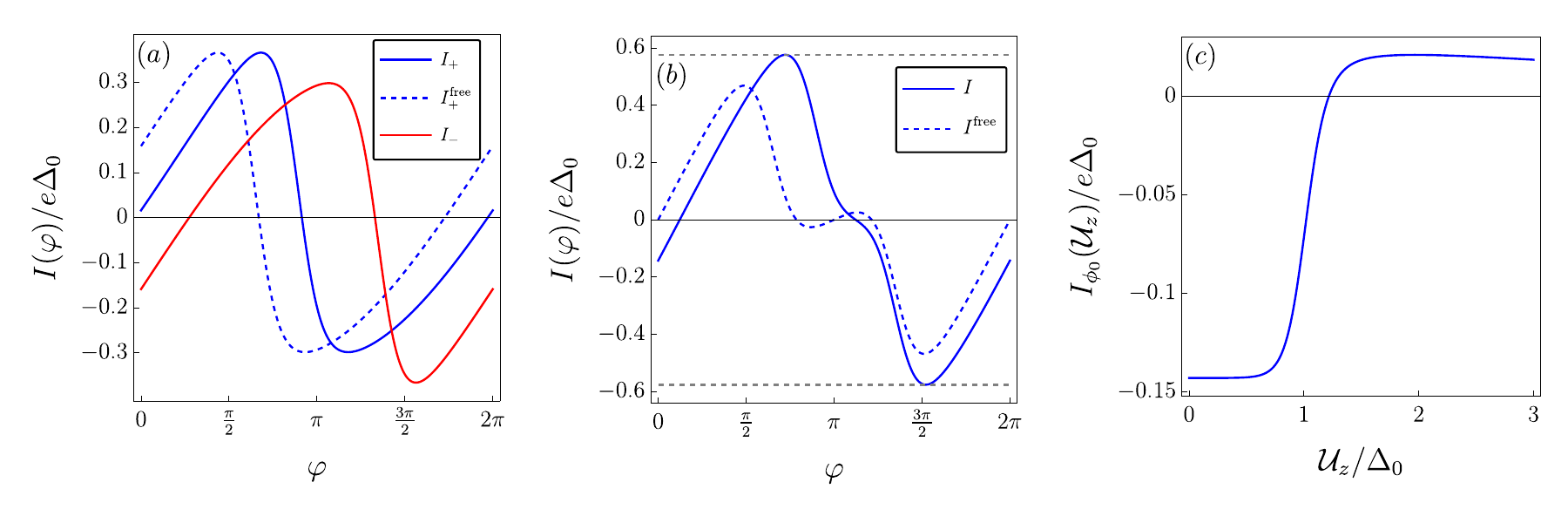}
    \caption{(a) Comparison of the current-phase relation for the lower edge ($I_-$), upper edge without the tip ($I^{\text{free}}_+$), and upper edge in presence of the tip ($I_+$), with $\mathcal{U}_z/\Delta_0=0.5$. (b) Comparison of the current-phase relation for the full system (both edges) in absence ($I^{{\rm free}}$) and presence ($I$) of the tip, with $\mathcal{U}_z/\Delta_0=0.5$. The dashed gray lines emphasize the lack of SDE in the total current. (c) Anomalous supercurrent $I_{\phi_0}=I(\varphi=0)$ as a function of $\mathcal{U}_z/\Delta_0$. In all plots, the field of the tip  (when present) is set to $\delta\mathcal{U}_z=\SI{5.4e-4}{\electronvolt}$ which, assuming $g=11.5$, corresponds to a magnetic field of $\SI{0.83}{\tesla}$. The temperature is $T=\SI{10}{\milli\kelvin}$.}
    \label{fig:Bz}
\end{figure*}

A first result of the present article is that for finite $\delta\mu$, the Josephson current remains identical to the one obtained in the absence of the tip. This fact closely resembles the Klein tunneling~\cite{Haldane_2017, Prada_2020}.
 Interestingly, this effect does not rely on the precise form chosen for $\delta\mu$, but it holds true for any spatially varying profile. Our result hence implies that the results of Ref.~\cite{Dolcini_2015} are robust with respect to chemical potential fluctuations.
From a mathematical point of view, the phenomenon emerges from the analytical form of $S_{\text{bound}_+}$, which reduces to 
\begin{equation}
    S_{\text{bound}_+}=S_{A_+}S_{\text{prop}_+}^{\delta\mu}.
\end{equation}
The Andreev scattering matrix $S_{A_+}$ has the form
\begin{equation}
    S_{A_+} =
    \begin{pmatrix}
        \textbf{0}&\sigma_{A_+}\\
        \sigma_{A_+}&\textbf{0}
    \end{pmatrix},
\end{equation}
while $S_{\text{prop}_+}^{\delta\mu}$ factorizes as
\begin{equation}
    S_{\text{prop}_+}^{\delta\mu}=
    \begin{pmatrix}
        e^{i\delta\mu \,d/v_F}\,\mathbb{I}_{2\times2} & \textbf{0}\\
        \textbf{0}&e^{-i\delta\mu \,d/v_F}\,\mathbb{I}_{2\times2}
    \end{pmatrix}
    S_{\text{prop}_+}^{\text{free}},
\end{equation}
with $S_{\text{prop}_+}^{\text{free}}$ a diagonal matrix, describing the propagation along the upper edge in the absence of the tip. Using the \textit{folding identity} for block matrices~\cite{Beenakker_1991, Beenakker_1992-I}, one can show that the contributions due to $\delta\mu$ cancel out in the computation of $D_+(z;\varphi)$ of Eq.~\eqref{eq:supercurrent_def}, thus leaving the current unchanged with respect to the unperturbed case.

The effect related to a finite $\delta\mathcal{U}_z$, {\it i.e.}, a magnetic tip with magnetization parallel to the edge spin polarization, is different with respect to the case just discussed. It turns out that the perturbation is again a modification of the propagation matrix $S_{\text{prop}_{+}}$
\begin{equation}
    S_{\text{prop}_{+}}^{\delta\mathcal{U}_z}=
    \begin{pmatrix}
        e^{i\delta\mathcal{U}_z \,d/v_F} &0&0&0\\
        0&e^{-i\delta\mathcal{U}_z \,d/v_F}&0&0\\
        0&0&e^{i\delta\mathcal{U}_z \,d/v_F}&0\\
        0&0&0&e^{-i\delta\mathcal{U}_z \,d/v_F}
    \end{pmatrix}
    S_{\text{prop}_+}^{\text{free}},
\end{equation}
but in a way that maintains the presence of the perturbation in Eq.~\eqref{eq:supercurrent_def}, where it manifests itself as a current-phase shift. Analytically, the phase shift is described by the following relation
\begin{equation}
    I_{+}(\varphi)\equiv I_{+}^{\text{free}}(\varphi-2\delta\mathcal{U}_z d/v_F)=-I_{-}(2\delta\mathcal{U}_z d/v_F-\varphi),
    \label{eq:shift}
\end{equation}
where $I_{+}^{\text{free}}(\varphi)$ is the CPR that one would obtain for the upper edge in absence of the tip. This fact is shown in Fig.~\ref{fig:Bz}(a), where the dashed curve represents the CPR in absence of the tip and the solid blue curve its translation due to the presence of the local $\delta\mathcal{U}_z$. For the parameters chosen ($\mathcal{U}_z/\Delta_0=0.5$) the shift in the CPR is such that the anomalous current in the upper edge almost vanishes.

From the point of view of non-reciprocal transport, a finite $\delta\mathcal{U}_z$ surprisingly does not result in any rectification, although the critical current is affected. Indeed, this is visible in Fig.~\ref{fig:Bz}(b): the maximum and minimum supercurrent amplitudes are equal, and thus no rectification is expected.
Finally, from Eq.~\eqref{eq:shift} one can see that the perturbation shifts the CPR of the perturbed edge, hence producing a finite anomalous supercurrent in the full structure, as shown in Fig.~\ref{fig:Bz}(c). This fact is remarkable, since it allows to manipulate the anomalous supercurrent by means of local perturbations. The sharp variation at $\mathcal{U}_z/\Delta_0=1$ is due to a quantum phase transition to a nodal phase occurring in the proximitized part. As a final remark, it is worth to mention that the concomitant presence of $ \mathcal{H}_t^{(1)}(x)$ and $\mathcal{H}_t^{(2)}(x)$ would lead exactly to the same results obtained for $ \mathcal{H}_t^{(2)}(x)$ alone.

\begin{figure*}[hbt]
    \centering
    \includegraphics[width=\linewidth]{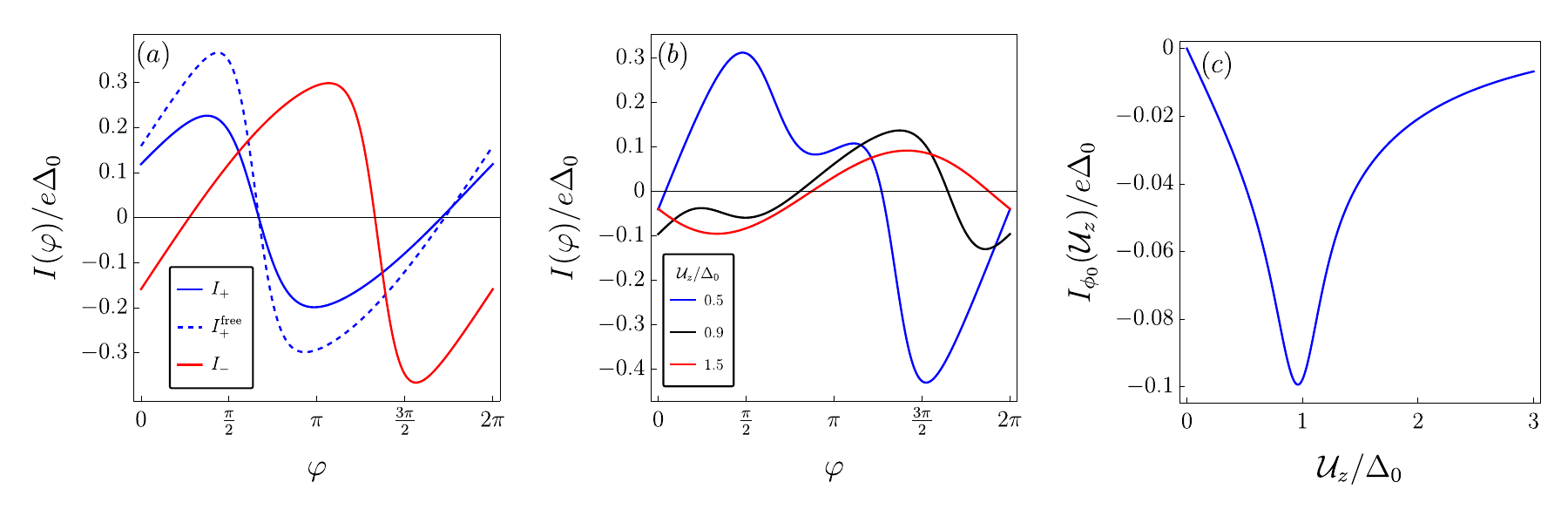}
    \caption{(a) Comparison of the current-phase relation for the lower edge ($I_-$), upper edge without the tip ($I^{\text{free}}_+$), and upper edge in presence of the tip ($I_+$), with $\mathcal{U}_z/\Delta_0=0.5$. (b) Comparison of the current-phase relations for the full system in presence of the tip, for different values of $\mathcal{U}_z/\Delta_0$, indicated in the legend. (c) Anomalous supercurrent $I_{\phi_0}$ as a function of $\mathcal{U}_z/\Delta_0$. In all  plots, the field of the tip  (when present) is set to $\delta\mathcal{U}_x=\SI{4.5e-4}{\electronvolt}$ which, assuming $g=11.5$, corresponds to a magnetic field of $\SI{0.69}{\tesla}$. The temperature is $T=\SI{10}{\milli\kelvin}$.
}
    \label{fig:Bx}
\end{figure*}

In the case of finite $\delta\mathcal{U}_x$, {\it i.e.}, a magnetic tip with perpendicular magnetization with respect to the edge spin polarization, the scenario is more complex. Indeed, such term acts as a barrier to the supercurrent, so that Majorana zero modes are localized\cite{MD1} for 
$\mathcal{U}_x<\Delta_0$, and the critical current in the perturbed edge decreases. This fact implies, for the full system, both the presence of SDE and of anomalous supercurrent.
This phenomenology arises from the fact that in the case of a transverse magnetic field, the perturbation does not commute with the free Hamiltonian. Therefore, in the perturbed region an energy gap is opened in the linear dispersion relations, causing the energy eigenstates in this magnetic gap to be exponentially suppressed. For this reason, these states contribute to the supercurrent in a minor way. Moreover, the transfer matrix $S_{\text{prop}_{+}}^{\delta\mathcal{U}_x}$ presents off-diagonal terms, that contain the spin-flip amplitudes (see the SM). The consequence of these effects is a change of the CPR and in particular a suppression in the magnitude  (Fig.~\ref{fig:Bx}(a)). This leads to an imbalance between the maximum and minimum value of the CPR in the two edges and, consequently, to a rectification that depends on the two magnetic scales present ($\mathcal{U}_z$ and $\delta\mathcal{U}_x$). In particular, if $\mathcal{U}_z$ is increased over $\Delta_0$, the diode effect gradually vanishes (cf. Fig.~\ref{fig:Bx}(b)). The amplitude imbalance manifests also at $\varphi = 0$ and so in the total anomalous supercurrent. This is represented in Fig.~\ref{fig:Bx}(c), where the anomalous supercurrent $I_{\phi_0}$ is plotted as a function of $\mathcal{U}_z/\Delta_0$: this quantity vanishes both for $\mathcal{U}_z\ll \Delta_0$ and $\mathcal{U}_z\gg \Delta_0$, reaching the maximum amplitude around $\mathcal{U}_z\sim \Delta_0$.

\begin{figure*}[hbt]
    \centering
    \includegraphics[width=\linewidth]{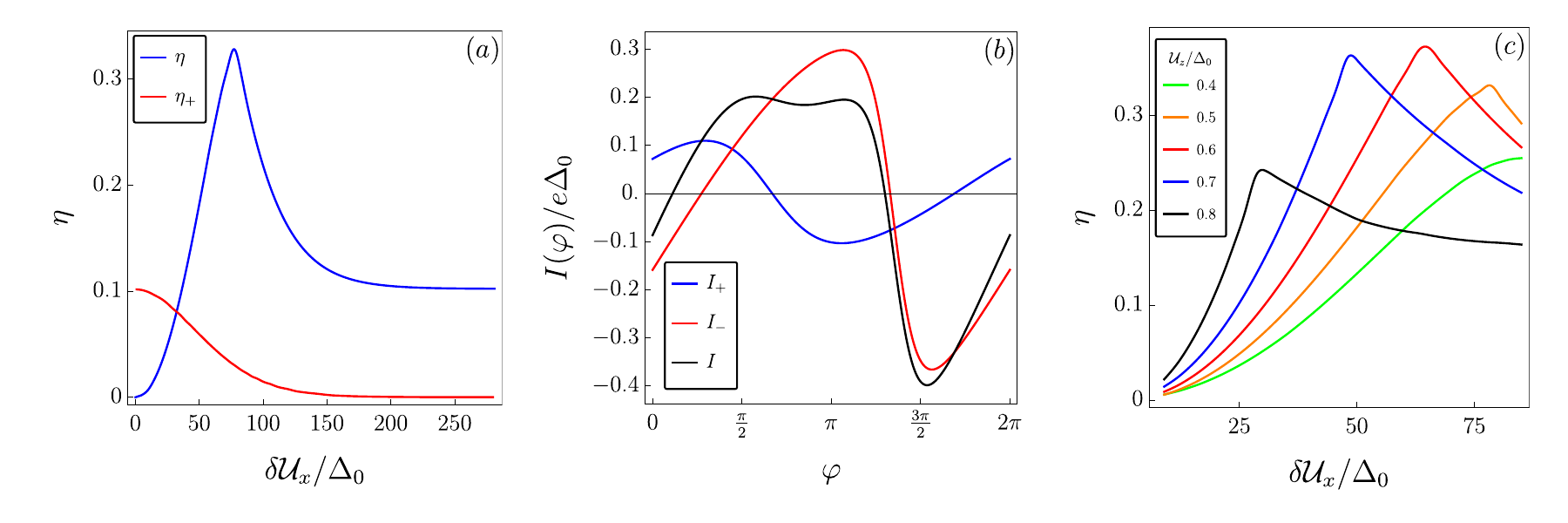}
    \caption{(a) Comparison between the rectification coefficient associated to the total current ($\eta(I)\equiv \eta$) and the one associated to the upper edge current ($\eta(I_+)\equiv \eta_+$) as a function of the tip field strength $\delta\mathcal{U}_x$, for $\mathcal{U}_z/\Delta_0=0.5$. $\eta$ presents a sharp peak around $\delta\mathcal{U}_x/\Delta_0 \sim 80$. (b) Comparison of the current-phase relation for the lower edge ($I_-$), upper edge ($I_+$), and full system ($I$) in presence of the tip with $\delta\mathcal{U}_x/\Delta_0 = 80$, obtained for $\mathcal{U}_z/\Delta_0=0.5$.
    (c) Comparison of the full system rectification coefficient $\eta$ for different values of the magnetic field $\mathcal{U}_z$.
    }
    \label{fig:DiodeEta}
\end{figure*}

In order to quantify the rectification effect, we compute the \emph{rectification coefficient}~\cite{Davydova_2022, Turini_2022, Strambini_2022}, which is defined as
\begin{equation}
    \eta[I] = \left|\frac{|\mathrm{Max}_\varphi\left\{I(\varphi)\right\}|- |\mathrm{Min}_\varphi\left\{I(\varphi)\right\}|}{|\mathrm{Max}_\varphi\left\{I(\varphi)\right\}|+ |\mathrm{Min}_\varphi\left\{I(\varphi)\right\}|}\right|.
    \label{eq:eta}
\end{equation}
Here, differently from what happens in more complex systems~\cite{signchange}, the presence of the outer absolute value in Eq.~(\ref{eq:eta}) does not mask non-trivial sign changes. In Fig.~\ref{fig:DiodeEta}(a) we plot the rectification coefficient for the total current ($\eta[I]\equiv\eta$), and for the upper edge current ($\eta[I_+]\equiv\eta_+$), as a function of the tip strength $\delta\mathcal{U}_x$. By increasing $\delta\mathcal{U}_x$, we find that $\eta$ reaches a (sharp) maximum and then decreases to a plateau.
Looking at the rectification of the single upper edge, it presents a simpler (monotonous) structure, with a vanishing $\eta_+$ at large $\delta\mathcal{U}_x$. 

To understand the physical origin of the peak in the rectification coefficient, in Fig.~\ref{fig:DiodeEta}(b) we plot the CPRs for the lower edge, upper edge (in presence of the tip) and for the full system, with $\delta\mathcal{U}_x$ set to the value for which $\eta$ reaches its maximum ($\equiv\delta\mathcal{U}_x^{\text{peak}}$). By comparing the three curves, one can see that the minimum of $I$ almost coincides with the minimum of $I_-$, since $I_+$ has a node close to where the minimum of $I_-$ occurs. On the other hand, due to the fact that $I_+$ is suppressed by the transverse magnetic field of the tip, the maximum of $I$ (which for $\delta\mathcal{U}_x=0$ coincides with the maximum of $I_+$) moves in the direction of the maximum of $I_-$ as $\delta\mathcal{U}_x$ is increased. In addition, for intermediate values of $\delta\mathcal{U}_x$ ({\it i.e.}, as long as $I_+$ is not completely suppressed) the maximum value of $I$ decreases in magnitude. For this reason there is an increasing unbalance between the maximum and minimum value of $I$.  For higher values of $\delta\mathcal{U}_x^{\text{peak}}$, the suppression of $I_+$ causes the maximum value of $I_-$ to be increasingly influential in the CPR, and consequently the maximum of $I$ begins to increase its value until, when $I_+$ is fully suppressed, it reaches the maximum of $I_-$.  This explains the presence of a peak in $\eta$ as well as the successive plateau. Indeed, for $\delta\mathcal{U}_x\gg\delta\mathcal{U}_x^{\text{peak}}$ the supercurrent in the upper edge is fully suppressed ($I_+\approx 0$) and therefore $I\approx I_-$. Thus, $\eta(\delta\mathcal{U}_x\to \infty) \approx \eta[I_-]=\eta_+(0)$, meaning that the plateau in $\eta$ corresponds to the value of the rectification for a single (unperturbed) edge.

 Given the mechanism leading to the non-monotonous behaviour in the rectification just described, one may wonder if by varying the field $\mathcal{U}_z$ (and so shifting the CPRs of both the upper and lower edge in opposite directions in phase space) it is possible to reach the peak at lower (experimentally more accessible~\footnote{Notice that for the value $\mathcal{U}_z = 0.5\Delta_0$ chosen in Fig.~\ref{fig:DiodeEta}(a-b) the magnetic field at the peak position $\delta\mathcal{U}_x\sim 80 \Delta_0$  is of the order of $\sim \SI{1.25}{\tesla}$.}) values of $\delta\mathcal{U}_x$. It turns out that this is indeed the case. In Fig.~\ref{fig:DiodeEta}(c) we report plots of $\eta$ as a function of $\delta\mathcal{U}_x$ for different values of $\mathcal{U}_z$: one can clearly see that for values of $\mathcal{U}_z$ higher then $\Delta_0/2$ (but still lower than $\Delta_0$), the peak moves to significantly smaller values of $\delta\mathcal{U}_x$, while maintaining sizeable (even larger) height. Finally, it is worth noting that in the case of simultaneous presence of $ \mathcal{H}_t^{(1)}(x)$ and $ \mathcal{H}_t^{(3)}(x)$, the results are qualitatively the same as in the case of $ \mathcal{H}_t^{(3)}(x)$ alone, due to the factorization of the scattering matrix.
 
In conclusion, we have considered the effects of local perturbations provided by a tip on the anomalous supercurrent and on the SDE in Josephson junctions based on a two dimensional topological insulator. We have shown that a capacitive coupling modelled as a chemical potential step has no effect on the transport properties, providing a generalization of the Klein paradox. Subsequently, we have shown that a magnetic coupling in the direction of the spin-momentum locking rigidly shifts, at the level of the single edge, the current-phase relation. It hence enables the manipulation of the anomalous supercurrent at the level of the full structure with two edges. However, it does not generate any SDE, although it influences the critical current. Finally, we have shown that a magnetic coupling perpendicular to the direction of spin-momentum locking is able to generate both anomalous supercurrent and SDE at the level of the full structure.

We argue that our results could be observable in InAs/GaSb quantum wells in the helical regime. Indeed, the numerical values considered for $v_F, \ g, \ \Delta_0$ are compatible with those actually measured in the proposed material~\cite{Kouwenhoven_2015, Rui-Rui_2016}, and magnetic fields of the tip below $\SI{1}{\tesla}$ are now experimentally achievable.\cite{Michlmayr2006, Phark2017, Haze2019} Even more favorable conditions in terms of $g$-factor and Fermi velocity are met in 2D spin-orbit coupled quantum gases, for which single edge theory could be effectively employed. Our work could hence provide a theoretical support to experiments aiming at the design of manipulable anomalous supercurrents and SDE.

\section*{Supplementary Materials}
The explicit expressions of the scattering matrices employed in this work are reported in the Supplementary Material.

\begin{acknowledgments}
N. T. Z. acknowledges the funding through the NextGeneration EU Curiosity-driven project ``Understanding even-odd criticality''.
The authors acknowledge the support from the project PRIN2022 2022-PH852L(PE3) TopoFlags - ``Non reciprocal supercurrent and topological transition in hybrid Nb-InSb nanoflags'' funded by the European community - Next Generation EU within the programme ``PNRR Missione 4 - Componente 2 - Investimento 1.1 Fondo per il Programma Nazionale di Ricerca e Progetti di Rilevante Interesse Nazionale (PRIN)''. S. H. acknowledges partial support by PNRR MUR Project No. PE0000023-NQSTI.
\end{acknowledgments}

\section*{Data Availability Statement}
The data that support the findings of this study are available from the corresponding author upon reasonable request.


\input{main.bbl}

\end{document}


\title{Supplementary Material for\\
``Anomalous supercurrent and diode effect in locally perturbed topological Josephson junctions''}
\author{Samuele Fracassi}
\affiliation{Dipartimento di Fisica, Universit\`a di Genova, Via Dodecaneso 33, 16146, Genova, Italy}
\author{Simone Traverso}
\thanks{Author to whom correspondence should be addressed: simone.traverso@edu.unige.it}
\affiliation{Dipartimento di Fisica, Universit\`a di Genova, Via Dodecaneso 33, 16146, Genova, Italy}
\author{Niccolo Traverso Ziani}
\affiliation{Dipartimento di Fisica, Universit\`a di Genova, Via Dodecaneso 33, 16146, Genova, Italy}
\affiliation{CNR-SPIN, Via Dodecaneso 33, 16146, Genova, Italy}
\author{Matteo Carrega}
\affiliation{CNR-SPIN, Via Dodecaneso 33, 16146, Genova, Italy}
\author{Stefan Heun}
\affiliation{NEST, Istituto Nanoscienze-CNR and Scuola Normale Superiore, Piazza San Silvestro 12, 56127 Pisa, Italy}
\author{Maura Sassetti}
\affiliation{Dipartimento di Fisica, Universit\`a di Genova, Via Dodecaneso 33, 16146, Genova, Italy}
\affiliation{CNR-SPIN, Via Dodecaneso 33, 16146, Genova, Italy}
\date{\today}

\maketitle    

\section{Omitted analytical expressions}
Here we report some explicit analytical expression omitted in the main text.\\
The scattering matrix describing the propagation on the upper edge in absence of the tip $ S_{\text{prop}_+}^{\text{free}}$ (Eq. 11) has the following form
\begin{equation}
    S_{\text{prop}_+}^{\text{free}} = \text{diag}[e^{ik_{e\uparrow}L},e^{-ik_{e\downarrow}L},e^{-ik_{h\downarrow}L},e^{ik_{h\uparrow}L}],
\end{equation}
where
\begin{equation}
    k_{e\uparrow}=E+\mu+\mathcal{U}_z,
\end{equation}
\begin{equation}
    k_{e\downarrow} = -E-\mu+\mathcal{U}_z,
\end{equation}
\begin{equation}
    k_{h\downarrow}=-E+\mu-\mathcal{U}_z,
\end{equation}
\begin{equation}
    k_{h\uparrow}=E-\mu-\mathcal{U}_z.
\end{equation}
The block matrices entering in the scattering expression describing Andreev process (Eq. 10) are
\begin{equation}
    \sigma_{A+} =\begin{pmatrix}
    0 & e^{i\,\varphi/2-i\,\textup{arctan}\left[\frac{\sqrt{\Delta_0^2-(E+\mathcal{U}_z)^2}}{E+\mathcal{U}_z}\right]}\\
   e^{-i\,\varphi/2-i\,\textup{arctan}\left[\frac{\sqrt{\Delta_0^2-(E-\mathcal{U}_z)^2}}{E-\mathcal{U}_z}\right]} & 0
    \end{pmatrix}
\end{equation}
The \textit{folding identity} is \begin{equation}
    \textup{Det}\begin{pmatrix}
    a&b\\c&d
\end{pmatrix} = \textup{Det}(ad-aca^{-1}b),
\end{equation}
which holds for generic $2\times2$ matrices $a,b,c,d $ with $\textup{det}\,a \neq 0$. For the $\delta\mathcal{U}_x$ perturbation,  $S^{\delta\mathcal{U}_x}_{\text{prop}_+}$ has this form
\begin{equation}
    S^{\delta\mathcal{U}_x}_{\text{prop}_+} = S_{P2+}S_{M+}S_{P1+},
\end{equation}
where $S_{P1+}$ and $S_{P2+}$ are the free propagation matrices from the superconductors to the perturbation and from the perturbation to the superconductors, respectively.
\newpage
The two propagation matrix are
\[
    S_{P1+} = \begin{pmatrix}
        e^{ik_{e\uparrow}(x_0-d/2+L/2)}&0&0&0\\
        0&e^{ik_{e\downarrow}(x_0+d/2-L/2)}&0&0\\
        0&0&e^{ik_{h\downarrow}(x_0+d/2-L/2)}&0\\
        0&0&0&e^{ik_{h\uparrow}(x_0-d/2+L/2)}
    \end{pmatrix}
\]
\[
    S_{P2+} = \begin{pmatrix}
        e^{ik_{e\uparrow}(-x_0-d/2+L/2)}&0&0&0\\
        0&e^{ik_{e\downarrow}(-x_0+d/2-L/2)}&0&0\\
        0&0&e^{ik_{h\downarrow}(-x_0+d/2-L/2)}&0\\
        0&0&0&e^{ik_{h\uparrow}(-x_0-d/2+L/2)}
    \end{pmatrix}
\]

Finally $S_{M+}$ is the magnetic scattering matrix.

\begin{equation}
    S_{M+} = \begin{pmatrix}
        \sigma_{e+}&\mathbf{0}\\
        \mathbf{0}&\sigma_{h+}
    \end{pmatrix}
\end{equation}
where
\begin{equation}\sigma_{e+} = \begin{pmatrix}
    \frac{e^{i\,\mathcal{U}_zd}\sqrt{(E+\mu)^2-\delta\mathcal{U}_x^2}}{D_e}&-\frac{i\delta\mathcal{U}_x\,\text{sin}\,[d\sqrt{(E+\mu)^2-\delta\mathcal{U}_x^2}]}{D_e}\\
    -\frac{i\delta\mathcal{U}_x\,\text{sin}\,[d\sqrt{(E+\mu)^2-\delta\mathcal{U}_x^2}]}{D_e}&\frac{e^{-i\,\mathcal{U}_zd}\sqrt{(E+\mu)^2-\delta\mathcal{U}_x^2}}{D_e},
\end{pmatrix}\end{equation}
\begin{equation}
    \sigma_{h+} = \begin{pmatrix}
    \frac{e^{i\,\mathcal{U}_zd}\sqrt{(E-\mu)^2-\delta\mathcal{U}_x^2}}{D_h}&-\frac{i\delta\mathcal{U}_x\,\text{sin}\,[d\sqrt{(E-\mu)^2-\delta\mathcal{U}_x^2}]}{D_h}\\
    -\frac{i\delta\mathcal{U}_x\,\text{sin}\,[d\sqrt{(E-\mu)^2-\delta\mathcal{U}_x^2}]}{D_h}&\frac{e^{-i\,\mathcal{U}_zd}\sqrt{(E-\mu)^2-\delta\mathcal{U}_x^2}}{D_h}
    \end{pmatrix}
\end{equation}
with
\[
    D_e = \sqrt{(E+\mu)^2-\delta\mathcal{U}_x^2}\,\text{cos}\,[d\sqrt{(E+\mu)^2-\delta\mathcal{U}_x^2}]-i(E+\mu)\,\text{sin}\,[d\sqrt{(E+\mu)^2-\delta\mathcal{U}_x^2}]
\]
\[
    D_h = \sqrt{(E-\mu)^2-\delta\mathcal{U}_x^2}\,\text{cos}\,[d\sqrt{(E-\mu)^2-\delta\mathcal{U}_x^2}]-i(E-\mu)\,\text{sin}\,[d\sqrt{(E-\mu)^2-\delta\mathcal{U}_x^2}]
\]
In order to get the scattering matrices of the lower edge it is necessary to perform the substitution $\mathcal{U}_z\rightarrow-\mathcal{U}_z$. This is a consequence of the fact that the Zeeman term is the only one that distinguishes the chirality of the edges.

%% file: main.bbl
%